\documentclass{article}

\usepackage{amsmath} 
\usepackage{amssymb} 
\usepackage[mathcal,mathscr]{eucal} 
\usepackage{eufrak} 
\usepackage{epsfig} 
\usepackage{amsthm} 
\usepackage{graphicx} 
\usepackage[numbers, sort&compress]{natbib} 
 
\begin{document} 
 
%%%%%%%%%%%%%%%%%%%%%%%% 
%%%%%   TITLE    %%%%%%% 
%%%%%%%%%%%%%%%%%%%%%%%% 
 
\begin{center}
\Large{\textbf{On nonlinear evolution and supraluminal communication between finite quantum systems}}
\end{center}

\begin{center} 
M.\ Ferrero\\
Dpto.\ F\'{\i}sica, Universidad de Oviedo, Spain\\
\texttt{maferrero@uniovi.es}

\medskip

D. Salgado \& J.L. S\'{a}nchez-G\'{o}mez\\ 
Dpto.  F\'{\i}sica Te\'{o}rica, Universidad Aut\'{o}noma de Madrid, Spain \\
\texttt{david.salgado@uam.es} \& \texttt{jl.sanchezgomez@uam.es} \\ 
\end{center}

\medskip

\vspace{1cm}

\begin{center}
\begin{minipage}{10cm} 
We revise the 'no-signaling' condition for the supraluminal communication between two spatially separated finite quantum systems of \emph{arbitrary} dimensions, thus generalizing a similar preceding approach for two-qubits:  non-linear evolution does not necessarily imply the possibility of supraluminal communication between any sort of finite quantum systems.
\end{minipage} 
\end{center}

\vspace{1cm}

%%%%%%%%%%%%%%%%%%%%%%%%%%%%%%%%%%%%%%%%%%%%%%%%%%%%%%%%%%%%
% The main text of your paper   begins here                          %
%%%%%%%%%%%%%%%%%%%%%%%%%%%%%%%%%%%%%%%%%%%%%%%%%%%%%%%%%%%%

\section{Introduction}%SECTION 1 HEADING TYPE HERE}
%TYPE YOUR TEXT HERE FOR SECTION 1. 
Though up to now there is no experimental indication why quantum evolution may be nonlinear, it has been traditionally considered both as a possible way out to the measurement problem\cite{Wig62a} or as matter of theoretical considerations to be contrasted with high-finesse experiments\cite{Wei89a}. One of the most remarkable consequences of these considerations\cite{Gis90a} was the possibility, under the nonlinearity assumption, of supraluminal communication between two spatially separated parties. This soon led some authors to conclude that any nonlinear quantum evolution would necessarily entail the possibility of such a communication\cite{Gis89a,GisRig95a} and even to consider the relativistic postulate of 'no-faster-than-light' phenomena as the theoretical basis for the quantum evolution to be linear\cite{SimBuzGis01a}. Recently\cite{FerSalSan03b} we have proven that this implicaction is not strict, i.e.\ that there exist possible nonlinear quantum evolutions not implying this fatal supraluminal communication.\\

Here we extend our previous result to finite quantum systems of arbitrary dimensions. We formulate the 'no-signaling' condition for these systems and show a full-flegded infinity of examples fulfilling this condition. Everything is expressed in Bloch space language\cite{Kim03a,ByrKha03a}, i.e.\ the states of quantum systems are expressed as 

\begin{equation}
\rho(t)=\frac{1}{N}\left(\mathbb{I}_{N}+\mathbf{r}(t)\cdot\mathbf{\sigma}\right)
\end{equation}

\noindent and orthogonal projectors as

\begin{equation}
P=P_{0}\mathbb{I}_{N}+\mathbf{P}\cdot\mathbf{\sigma}
\end{equation}
\noindent where $\mathbf{r}(t)$ is a time-dependent so-called Bloch vector belonging to a particular convex subset of $\mathbb{R}^{N^{2}-1}$, $\mathbf{\sigma}\equiv\left(\sigma_{1},\dots,\sigma_{N^{2}-1}\right)$ are the traceless orthogonal generators of $SU(N)$ and $(P_{0},\mathbf{P})\equiv(P_{0},P_{1},\dots,P_{N^{2}-1})$ are real numbers subjected to certain restrictions (cf.\ \cite{Kim03a} for the details).

\section{The 'no-signaling' condition}

%TYPE YOUR TEXT HERE FOR SECTION 2. 

As remarked in \cite{FerSalSan03b}, the impossibility of communication through the projection postulate, i.e.\ at a speed faster than that of light, is obtained only after imposing that \emph{the \textbf{probability distribution} of any observable of one subsystem \textbf{only} depends on its own reduced state}. The mathematical translation of this criterion is straightforward provided one is familiar with the preceding language. Let us consider a two-partite system of subsystems $1$ and $2$, which have dimensions $N_{1}$ and $N_2$, respectively. Their common density matrix, using a tensor product basis, will be given by

\begin{equation}
\rho_{12}=\frac{1}{N_{1}N_{2}}\left(\mathbb{I}_{N_{1}N_{2}}+\mathbf{r}^{(1)}\cdot\mathbf{\sigma}\otimes\mathbb{I}_{N_{2}}+\mathbb{I}_{N_{1}}\otimes\mathbf{r}^{(2)}\cdot\lambda+\sum_{ij}r_{ij}^{(12)}\sigma_{i}\otimes\lambda_{j}\right)
\end{equation}

\noindent and an orthogonal projector for each of them by

\begin{equation}
P^{(1)}=P_{0}^{(1)}\mathbb{I}_{N_{1}}+\mathbf{P}^{(1)}\cdot\sigma\quad P^{(2)}=P_{0}^{(2)}\mathbb{I}_{N_{2}}+\mathbf{P}^{(2)}\cdot\lambda
\end{equation}

\noindent respectively, where $\sigma$ ($\lambda$) stands for the traceless orthogonal generators of $SU(N_{1})$ ($SU(N_{2})$) and $\mathbf{P}^{(1)}$ ($\mathbf{P}^{(2)})$ is a $(N^{2}_{1}-1)$($(N^{2}_{2}-1)$)-dimensional vector restricted to some given subset\footnote{Namely, $P_{0}=P_{0}^{2}+\mathbf{P}\cdot\mathbf{P}$ and $2P_{0}P_{n}+z_{ijn}P_{i}P_{j}=P_{n}$, where $z_{ijk}\equiv g_{ijk}+if_{ijk}$, the latter denoting the completely symmetric and antisymmetric tensors of the Lie algebra $\mathfrak{su}(N_{j})$, respectively.}.\\
Suppose now that an orthogonal projector $(u_{0},\mathbf{u})$ is measured upon subsystem $2$. Then $N_{2}$ possible outcomes $(u_{0}^{(k)},\mathbf{u}^{(k)})$ will result with probabilites $p_{k}=u_{0}^{(k)}+\mathbf{u}^{(k)}\cdot\mathbf{r}^{(2)}$ given by the trace rule. Also, the projection postulate allows us to conclude that after such a measurement, the reduced density operator for its partner, subsystem $1$ will be given by

\begin{equation}
\rho_{k}^{(1)}(0)=\frac{1}{N_{1}}\left(\mathbb{I}_{N_{1}}+\mathbf{r}^{(1;k)}\cdot\sigma\right)
\end{equation} 

\noindent where $\mathbf{r}^{(1;k)}$ is an $(N^{2}_{1}-1)$-dimensional vector ($k=1,\dots,N_{2}$ possible outcomes) dependent on the joint state $\mathbf{r}^{(1)},\mathbf{r}^{(2)},r_{ij}^{(12)}$ and on the measured observable $(u_{0},\mathbf{u})$:

\begin{equation}
\mathbf{r}^{(1;k)}_{j}=\frac{u_{0}^{(k)}r_{j}^{(1)}+\sum_{n=1}^{N_{2}^{2}-1}r_{jn}^{(12)}u_{n}^{(k)}}{u_{0}^{(k)}+\mathbf{u}^{(k)}\cdot\mathbf{r}^{(2)}}\equiv r^{(1;k)}_{j}(0)
\end{equation}

In these conditions, the probability distribution $\mathbb{P}$ of an arbitrary orthogonal projector $(v_{0},\mathbf{v})$ with $p=1,\dots,N_{1}$ possible outcomes $(v_{0}^{(p)},\mathbf{v}^{(p)})$ at time $t$ of subsystem $1$ will be given by 

\begin{equation}
\mathbb{P}^{(1)}(t;v^{(p)})=\sum_{k=1}^{N_{2}}(u_{0}^{(k)}+\mathbf{r}^{(2)}\cdot\mathbf{u}^{(k)})(v_{0}^{(p)}+\mathbf{v}^{(p)}\cdot\mathbf{r}^{(1)}(t;\mathbf{r}^{(1;k)}(0))
\end{equation}

\noindent where $\mathbf{r}^{(1)}(t;\mathbf{r}^{(1;k)}(0))$ denotes the Bloch vector of subsystem $1$ at time $t$ with initial condition $\mathbf{r}^{(1;k)}(0)$.\\
 
The 'no-signaling' condition can then be easily formulated. The independece with respect to other partners' reduced state and their mutual correlations will be expressed as

\begin{eqnarray}\label{NoSig1}
\frac{\partial\mathbb{P}^{(1)}(t;v^{(p)})}{\partial r_{k}^{(2)}}&=&0\\
\label{NoSig2}\frac{\partial\mathbb{P}^{(1)}(t;v^{(p)})}{\partial r_{ij}^{(12)}}&=&0
\end{eqnarray}

\noindent Finally, the independence with respect to observables to be measured in spatially separated subsystems will be expressed as

\begin{equation}\label{NoSig3}
\frac{\partial\mathbb{P}^{(1)}(t;v^{(p)})}{\partial u^{(k)}_{\mu}}=0\quad\mu=0,1,\dots,N_{1}^{2}-1
\end{equation}

These three conditions are the mathematical translation of the previously formulated 'no-signaling' condition. The reader may check for himself that, as expected, the usual linear quantum evolution fulfills each of them (see also below).

\section{Consequences}

One of the main consequences of eqs.\ (\ref{NoSig1}), (\ref{NoSig2}) and (\ref{NoSig3}) arises after noticing that they must be valid for any particular value of the parameters involved, which implies $\mathbf{r}^{(i)}(t;\mathbf{r}_{k})=A^{(i)}(t)\mathbf{r}_{k}$, where $A^{(i)}(t)$ is a time-dependent matrix. In other words, the reduced dynamics in absence of interactions (spatial separation) must be linear. Note that this does not exhaust the possibility of having nonlinear joint evolution. Indeed reduced linearity in absence of interactions entails neither joint linearity nor even reduced unitarity. Expressing this in Bloch vector language, if $(\mathbf{r}^{(1)}(t),\mathbf{r}^{(2)}(t),r_{ij}^{(12)}(t))$ denotes the Bloch vector of a two-partite system and if $H=H_{0}\mathbb{I}_{N_{1}N_{2}}+\mathbf{H}\cdot\sigma_{12}$ ($\mathbf{H}=(\mathbf{H}^{(1)},\mathbf{H}^{(2)},H^{(12)})$ and $\sigma_{12}=(\sigma\otimes\mathbb{I}_{N_{2}},\mathbb{I}_{N_{1}}\otimes\lambda,\sigma\otimes\lambda)$) denotes its joint Hamiltonian, then any evolution given by

 \begin{eqnarray}
\mathbf{r}^{(1)}(t)&=&\mathbf{F}_{1}(t;H,\mathbf{r}(0))\\
\mathbf{r}^{(2)}(t)&=&\mathbf{F}_{2}(t;H,\mathbf{r}(0))\\
r^{(12)}(t)&=&F_{12}(t;H,\mathbf{r}(0))
\end{eqnarray}

\noindent such that in absence of interactions ($H^{(12)}=0$) satisfies

\begin{eqnarray}
\label{RedLin1}\mathbf{r}^{(1)}(t)&=&M^{(1)}(t;\mathbf{H}^{(1)})\mathbf{r}^{(1)}(0)\\
\label{RedLin2}\mathbf{r}^{(2)}(t)&=&M^{(2)}(t;\mathbf{H}^{(2)})\mathbf{r}^{(2)}(0)
\end{eqnarray}

\noindent where $M^{(k)}(t;\mathbf{H}^{(k)})$ denotes a time-dependent matrix depending only on the Hamiltonian of the $k$th subsystem, is free of supraluminal communication.\\

It should be clear that this nonlinearity only affects the evolution and never the static structure of the theory, i.e.\ the principle of superposition of quantum states at a given instant of time is still valid, only the evolution of these states is affected.\\

Alternatively, one can express these nonlinearities through the evolution equations:

\begin{eqnarray}
\frac{dr^{(1)}_{i}}{dt}&=&\left(\sum_{m,n=1}^{N_{1}^{2}-1}f_{imn}^{(1)}H^{(1)}_{m}r^{(1)}_{n}+\sum_{j,m,n=1}^{N_{1}^{2}-1}f_{ijm}^{(1)}H^{(12)}_{jn}r^{(12)}_{mn}\xi_{i;jn}^{(1)}(\mathbf{r}^{(1)},\mathbf{r}^{(2)},r^{(12)})\right)\nonumber\\
&&\\
\frac{dr^{(2)}_{i}}{dt}&=&\left(\sum_{m,n=1}^{N_{2}^{2}-1}f^{(2)}_{imn}H^{(2)}_{m}r^{(2)}_{n}+\sum_{j,m,n=1}^{N_{2}^{2}-1}f^{(2)}_{ijm}H^{(12)}_{jn}r^{(12)}_{nm}\xi_{i;jn}^{(2)}(\mathbf{r}^{(1)},\mathbf{r}^{(2)},r^{(12)})\right)\\
\frac{dr^{(12)}_{pq}}{dt}&=&2\left(\sum_{i,j=1}^{N_{1}^{2}-1}f^{(1)}_{jip}H^{(1)}_{j}r^{(12)}_{iq}+\sum_{i,j=1}^{N_{2}^{2}-1}f^{(2)}_{jip}H^{(1)}_{j}r^{(12)}_{qi}+\right.\nonumber\\
&+&\sum_{i,j=1}^{N_{1}^{2}-1}\sum_{m,n=1}^{N_{2}^{2}-1}\textrm{Im}\left[z_{ijp}^{(1)}z_{mnq}^{(2)}\right]H_{im}^{(12)}r_{jn}^{(12)}\xi_{pq;im}(\mathbf{r}^{(1)},\mathbf{r}^{(2)},r^{(12)})+\nonumber\\
&+&\left.\sum_{i,j=1}^{N_{1}^{2}-1}f_{ijp}^{(1)}H^{(12)}_{iq}r^{(1)}_{j}\xi_{pq;iq}^{(12)}(\mathbf{r}^{(1)},\mathbf{r}^{(2)},r^{(12)})+\sum_{i,j=1}^{N_{2}^{2}-1}f_{ijp}^{(1)}H^{(12)}_{qi}r^{(2)}_{j}\xi_{pq;iq}^{(12)}(\mathbf{r}^{(1)},\mathbf{r}^{(2)},r^{(12)})\right)\nonumber\\
\end{eqnarray}

\noindent where the functions $\xi$ are completely arbitrary. Notice that in absence of interactions ($H^{(12)}=0$), one recovers the usual well-known quantum evolution.

\section{Conclusions}%CONCLUSIONS SECTION HEADING TYPE HERE}

The main two conclusions to be drawn are that (i) nonlinear evolution does not necessarily imply the possibility of supraluminal communication between two arbitrary finite quantum systems, and (ii) non linear terms, in order to fulfill the no-signaling condition, must be necessarily associated to interactions.\\

This reopens a door, originally suggested by Wigner, to explore possible solutions to the measurement problem without contradicting other well contrasted theories.\\

A third generalization of this approach can be undertaken by focusing on non-projective measurements, but on generalized measurements, i.e.\ on POVM's \cite{Per93a}.

%%%%%%%%%%%%%%%%%%%%%%%%%%%%%%%%%%%%%%%%%%%%%%%%%%%%%%%%%%%%
% Doing Acknowledgement                                              %
%%%%%%%%%%%%%%%%%%%%%%%%%%%%%%%%%%%%%%%%%%%%%%%%%%%%%%%%%%%%

\section*{Acknowledgements}
We acknowlegde financial support from Spanish Ministry of Science and Techmology through project no.\ FIS2004-01576. M.F.\ also acknowledges financial support from Oviedo University (ref.\ no.\ MB-04-514).

%%%%%TYPE ACKNOWLEGDEMENTS HERE.

%%%%%%%%%%%%%%%%%%%%%%%%%%%%%%%%%%%%%%%%%%%%%%%%%%%%%%%%%%%%
% Doing Appendix(ices) - Appendix A & B are shown below    %  
%%%%%%%%%%%%%%%%%%%%%%%%%%%%%%%%%%%%%%%%%%%%%%%%%%%%%%%%%%%%

\appendix

%%%%%%%%%%%%%%%%%%%%%%%%%%%%%%%%%%%%%%%%%%%%%%%%%%%%%%%%%%%%
% Doing references:                                        %
%%%%%%%%%%%%%%%%%%%%%%%%%%%%%%%%%%%%%%%%%%%%%%%%%%%%%%%%%%%%

%\bibliographystyle{apsrev}
%\bibliography{/home/david/Bibliography/Biblio}

\end{document}